# The Syntax of the Accounting Language: A First Step

**Author:**

***Dr Frederico Botafogo**, Otago Business School (Te Kura Pakihi),
University of Otago, Dunedin, New Zealand*

**Contact Details:**

PO Box 56, Dunedin 9054, New Zealand

frederico.botafogo@otago.ac.nz

**Abstract:**

We review and interpret two basic propositions published by Ellerman (2014) in this journal. The propositions address the algebraic structure of T-accounts and double-entry bookkeeping (DEB). The paper builds on this previous contribution with the view of reconciling the two, apparently dichotomous, perspectives of accounting measurement: the one that focuses preferably on the stock of wealth and to the one that focuses preferably on the flow of income. The paper claims that T-accounts and DEB have an underlying algebraic structure suitable for approaching measurement from either or both perspectives. Accountants' preferences for stocks or flows can be framed in ways which are mutually consistent. The paper is a first step in addressing this consistency issue. It avoids the difficult mathematics of abstract algebra by applying the concept of syntax to accounting numbers such that the accounting procedure qualifies as a formal language with which accountants convey meaning.

# The Syntax of the Accounting Language: A First Step

## *1. Introduction*

Accounting measurement may be approached from two, presumably dichotomous, perspectives: The accountant measures wealth, which expresses the stocks of value, at several points in time and reports any differences over time as income, which expresses the flows of value within time intervals. Alternatively, by keeping track of transactions the accountant measures income, as a flow of value, and then reports the resulting wealth as the stock of value which is derived from up-dating the original wealth by the measured flow. The former approach is sometimes called the valuation approach, while the latter is the transactions approach to measurement (Willett, 1987). The FASB and the IASB conceptual frameworks, which support the prevailing standards for accounting worldwide, are rooted in the former approach, usually referred to as the 'asset/liability view' (Hales, Rees and Wilks, 2016). Since the alternative view, the 'revenue/expense view', goes against the standards, it has been 'exiled' from classrooms and university lecture halls, only being kept alive by a handful of scholars (e.g., Biondi, 2011; Dichev, 2016) who resist the conceptual frameworks on various grounds. This dichotomy presents the theorist with a foundational research problem, since the question of which approach is 'better'[1] remains an open question to this day.

In this paper, we do not seek a solution. Rather, our modest intention is to make the case for a better understanding of certain structural characteristics that pertain to accounting numbers. We claim that dealing with the mathematical structure which supports the accounting numbers helps with the framing of the problem. This is a first step in approaching the problem from a fresh perspective, such that in due course it may be tackled to the theorist's satisfaction.

---

[1] The criterion for 'better' varies greatly, depending on the theorist. For some examples, in no way comprehensive of what may be found in the literature, Bromwich, Macve, and Sunder (2010) claim there is a logical inconsistency with the way the conceptual frameworks define wealth; Biondi (2011), considering that firms have responsibilities beyond their shareholders and thus should not focus on creating wealth only, argues for the need to account for the flow of value across a firm's stakeholders.

Within this research context, we articulate the forthcoming discussion further to Ellerman's (2014), 'On double-entry bookkeeping: The mathematical treatment', published in this journal. In that paper, double-entry bookkeeping (DEB) is discussed from a structural perspective, where the term *structure* refers to the mathematics of abstract algebra. We build upon this contribution. Further, we have been instructed by the editor to proceed with an outline of our argument rather than a detailed discussion of abstract algebra. Thus, given the foregoing three points – the research context, the algebraic underpinning to articulate the problem, and the editor's constraint – we have organised our paper as follows. First, we explain the problem. Next, we present some useful background ideas. Then we review, interpret, and illustrate the two most basic propositions by Ellerman (2014). Finally, we discuss the problem and conclude.

## 2. *The accounting measurement problem*

At the most abstract level, modern accounting must address two foundational issues: First, how do entities combine economic resources (i.e., assets) and how do they combine stakeholders' demands (e.g., liabilities when the demands are legally enforcing[2])? Second, how do they keep track of economic processes (e.g., production) to evaluate performance?

From a measurement perspective, Solomons (1978) articulates the view that these issues should be addressed as 'financial mapmaking', a view later incorporated by the standard setters as a desired fundamental qualitative characteristic of financial statements under the label of 'representational faithfulness'. That view seems appropriate to address the first issue in which assets and liabilities are conceived as elements that exist in some underlying economic landscape. However, the same is not true when addressing the second issue in which the focus is on processes and time evolution.

---

[2] Demands by stakeholders other than the shareholders are not classified as liabilities. They are, however, subjected to intense investigations by accounting scholars (e.g., reporting that integrates the planet, people, and profits). Those demands are not (as yet) addressed by the financial standards because the applicable measures are clearly not financial. Nevertheless, they do characterise an accounting measurement problem.

Indeed, some elements in the map may not have been simultaneously observed, in which case the map is not a straightforward representation of reality but rather a construct that provides a narrative expected to be consistent with the observed phenomena. Further, on closer inspection, even the appropriateness of the mapmaking analogy for addressing the first issue comes into question when one realises that assets and liabilities are socially constructed in view of past history and future expectations. Both issues are inherently entangled and result in a plethora of problems plaguing the theory and the practice of accounting, as follows.

Table 1, cut and pasted from Moore (2009), provides a partial list of some significant problems in accounting measurement. To this list we add Hines (1988) and Chiapello (2008), who both address the problem of reflexivity[3] in accounting, and Beaver and Demski (1979), who claim that the very idea of measuring income does not apply in any real-world context, since it is only valid under the assumption of complete and perfect markets.[4]

Our own view is that tracking flows should take precedence over measuring stocks. To phrase this precisely, one should complete the fundamental measurement of flows first, then *derive* the measures applicable to stocks. This is informed by the importance of flows today.

> *In this world, growth increasingly depends on the ability of yourself, your community, your town, your factory, your school and your country to be connected to more and more of the* flows *of knowledge and investment – and not just rely on* stocks *of stuff.*
>
> *Over centuries, notes John Hagel, who currently co-heads Deloitte's Center for the Edge, business has 'been organized around stocks of knowledge as the basis for value creation. […] The challenge in a more rapidly changing world is that knowledge stocks depreciate at an accelerating rate. In this kind of world, the key source of economic value shifts from stocks to flows'.* (Friedman, 2019)

The precedence of flows being accepted, the question becomes how to deal with it. Further to Moore (2009), we suggest framing accounting measurement with the mathematics of quantum physics.

---

[3] In communicating reality, accountants construct reality (Hines, 1988). An accounting statement brings into being that of which it speaks, as when an auditor designates a firm to be solvent (Chiapello, 2008).

[4] They championed the now mainstream Information Content approach to accounting.

**TABLE 1**
**Problems with Accounting Measures**

| Problem | Short Description of Problem |
|---|---|
| (1) Measurement Scales: (Bierman 1969) | Nominal, ordinal, interval, and ratio scales: Prevalence of nonratio scales in practice limits the ability to assess distances between measures. |
| (2) Measurement Approaches: (Edwards and Bell 1961; Chambers 1966; Sweeney 1930) | No agreement on available approaches: (a) cost, with or without price-level adjustments and (b) value; (1) replacement value; (2) realizable value; (3) net realizable value; (4) discounted cash flows; and (5) value of similar productive capacity. |
| (3) Traditional Transactions Income versus Economic Income: (Hicks 1939; Mitchell 1967; Revsine 1970) | Transaction approach is a weak substitute for the opportunity costs hypothesized in pure economic theory. |
| (4) Incorrigibility of Allocations: (Thomas 1969; Devine 1985) | It is impossible to derive a single defensible basis for allocating costs. |
| (5) Transfer pricing: (*The Economist* 2007) | Transfer prices can move income from department to department—distorts tax and responsibility accounting systems. |
| (6) Market adjustments are often counterintuitive: (Brubaker 2008; Katz and Reason 2008) | Market adjustments for liabilities in companies with declining credit worthiness defy economic common sense. |
| (7) Circular effect and inability to audit instruments based on future market conditions: (Benston 2006) | Many types of financial instruments cannot be objectively valued on an ongoing basis; reported values may cause market changes. |
| (8) Management incentives for income manipulation: (Core et al. 2003) | Reward systems based on market or accounting performance are associated with opportunistic choice of accounting procedures. |

Quantum measurement is predicated on waves of probabilities, an inherently flow-oriented construct. Further, the language of quantum physics is linear algebra, at the very basis of which lies the concept of algebraic groups. The key contribution by Ellerman, summarised in his 2014 paper published in this journal, is to have realised that T-accounts satisfy the axioms of an algebraic group. We build on this understanding such that accounting by DEB may be framed in a mathematical language that supports a flow-oriented approach to measurement.

## 3. Background

Mathematics qualifies as a language, specifically as a formal language. The structural analysis of that language is framed by abstract algebra in an analogous fashion to the structural analyses of natural languages such as English or French being framed by syntax.[5] We open this section with an example that illustrates how syntax contributes to communication. A rigorous use of syntax prevents ambiguity, since words by themselves are not sufficient to convey meaning with clarity. For the sake of the forthcoming discussion, the point being made is that the same is true of mathematics where numbers by themselves are not sufficient to convey meaning.[6]

We then proceed with another example whereby waves are best described by means of imaginary numbers. For the sake of the discussion, the point is that a rich syntax supports semantics: meaning built-in within the language helps with interpreting real-world phenomena.

We conclude the section with the final example of Arabic numerals, whose decimal structure supports the algorithms for adding and multiplying numbers. In this case, the point is that semantics (i.e., meaning) may be embedded within the language's structure itself.

*3.1 Syntax – The structure of a language*

The sentence 'I saw a man on a hill with a telescope' admits five possible interpretations, as listed in table 2. The four initial interpretations illustrate the ambiguity of natural languages. The first interpretation is likely to be preferred by native English speakers because its ordering of words reflects the natural way in which meaning is conveyed in that language.[7] The fifth interpretation illustrates the ambiguity of context in natural languages, since 'saw' is not understood in this instance as the past simple of the verb 'see' but rather as the present tense of the verb 'saw', to cut or divide with a saw. Humans are quite good at perceiving context such that this interpretation is unlikely. However, the same is not true of computers.

---

[5] Syntax is the study of the formation of sentences and the relationship of their component parts ('Syntax', Encyclopaedia Britannica). It also investigates relations among sentences that are similar in order to identify their common *underlying structure*, as in 'John read the book' and 'The book was read by John'.

[6] See Roberts (1985) for measurement theory: a simple case occurs with nominal scales wherein assigning numbers to underlying objects is a valid procedure but then adding the assigned numbers is meaningless.

[7] There is an opposite symmetry in the case of numbers. For numbers to have their desired meaning, the order in which they are combined cannot affect how the combination is interpreted. In technical parlance, combination is framed by addition, which satisfies the axioms of commutativity and associativity.

As we are living in an era where all information, including specifically accounting information, is processed and even analysed by algorithms, a call for a clear understanding of the syntax applicable to accounting numbers should be readily welcome.

Abstract algebra is the branch of mathematics that deals with structures within which algorithms are devised and implemented. It provides the syntax for dealing with numbers.[8]

---

**TABLE 2**

**The importance of syntax in natural languages: an English sentence.**

## I saw a man on a hill with a telescope

Possible interpretations:

(i) There's a man on a hill, and I'm watching him with my telescope.

(ii) There's a man on a hill, whom I'm seeing, and he has a telescope.

(iii) There's a man, and he's on a hill that also has a telescope on it.

(iv) I'm on a hill, and I saw a man using a telescope.

(v) There's a man on a hill, and I'm sawing him with a telescope!

---

*3.2 Syntax to Semantics – The various ontological bases for a language*

The definition for the imaginary number, $i = \sqrt{-1}$, is nonsensical.[9] That is why it was originally called 'imaginary'. However, today *i* is understood to be as 'legit' a number as $\pi$ or 1 or 2. This is so because complex numbers, of which *i* is the archetypal example, are now embedded within the rules that define an algebraic field, which extends upon the algebraic group structure.[10]

---

[8] For an introductory discussion in the particular case of accounting, see Cruz Rambaud et al. (2010, ch.1).
[9] The square root is that number which when multiplied by itself yields the radicand. Since multiplying two positive numbers as well as multiplying two negative numbers yields a positive number in both cases, there should not be any square root of minus one.
[10] There is a hierarchy of algebraic structures in terms of increasing complexity: group, ring, field, vector spaces, tensor spaces, algebras and manifolds, etc. While the group structure typically supports addition, the field supports addition and multiplication and is thus the one necessary for basic arithmetic.

When articulating real and complex numbers within those rules, it is possible to derive from mathematical analysis only – that is, without the need to consider any real-world phenomenon – what is known as the Euler formula, namely $Ae^{i\omega t} = A\times\cos(\omega t) + A\times i\times\sin(\omega t)$. Since the trigonometric functions $\cos(\omega t)$ and $\sin(\omega t)$ provide straightforward representations for oscillatory phenomena, a connection is established between the 'imaginary' $i$ and waves.

Thus, a rich syntax supports semantics: an 'imaginary' object helps with interpreting and analysing the real-world phenomena of waves, such as observable sea waves or unobservable electromagnetic waves. Further, electromagnetic waves are not just unobservable; they have no ontological substance, since the medium wherein they would oscillate does not exist.[11] The only reason for one's referring to electromagnetic *waves* is because of the *i* in Euler's equation.

The foregoing is unlikely to be accepted unanimously. We failed to convince an anonymous referee of its relevance and validity within the accounting context. The referee wrote: 'The author(s) are fairly uncritical of Ellerman's […] (2014) approach, [which] like that of Nehmer et al.,[12] Mattesich,[13] and most others, tells us little about what the nature of assets, liabilities and equities are in terms of natural, non-accounting objects'. This suggests the epistemological approach whereby one considers the observable properties of objects or events, say sea waves, and then finds the mathematical objects best suited to represent them, the wave equation.

Here we are claiming that in addition to this approach, physics has demonstrated the validity of the inverse approach: mathematical elements conceived in the abstract help model objects and events that are unobservable at any single point in time, as is the case of assets and liabilities being constructed in view of past transactions and future expectations.

---

[11] This was shown by the Michelson-Morley experiment, called the most famous failed experiment in history, since it did not detect the existence of aether. Instead, it provided evidence in support of the theory of relativity.
[12] We believe the referee is referring to Cruz Rambaud, Pérez, Nehmer, and Robinson (2010).
[13] We believe the referee had in mind Mattessich (1957).

### *3.3 Semantics – Operationalising a language*

Semantics may be embedded within syntax. An easy-to-understand case is Arabic numerals, which are subjected to decimal placing such that the algorithms for addition and multiplication readily follow. We make sense of numbers through the practicing of these algorithms.

Further, there are instances of numbers (typically large numbers) that bear meaning only in terms of said algorithms. Anecdotally, time and again we come across students who confuse millions with billions because they fail to understand orders of magnitude. To remember that they differ by the factor $10^3$ is the key to understanding them.

Moreover, as Macbeth (2011, p.60) explains, one is able to think within the Arabic system in ways not available to Roman numerals, for example: '… collections of signs in [Arabic] mathematics do not merely record results; they actually embody the relevant bits of mathematical reasoning'. (highlights added)

It is important to understand how the structure underlying DEB can be operationalised, not just in terms of stocks but also in terms of flow. As an example, consider activity-based costing (ABC) which translates steps along a production process into cost objects that are then added and yield the final cost of a product. Today, advances in costing techniques are likely to come from implementing algorithms using AI. As a precondition, the software must be structured in ways consistent with DEB. This is the point being made here.

### 4. *Review and extension of two propositions by Ellerman (2014)*

There are two, most fundamental propositions embedded within Ellerman's (2014) discussion of DEB. We present and review them next.

The first proposition states that T-accounts satisfy the axioms of an algebraic group. In plain English, this means that T-accounts are conceptual objects that can be combined and when this occurs another T-account obtains. Combining cash, receivables, and inventory yields the current asset account. Combining sales, COGS, and expenses yields the net income account.

For measurement purposes, this adds a level of complexity which, to our knowledge, is not addressed in the existing literature. Indeed, classifying resources under a specific account title appears to be a nominal measurement process. The ability to combine titles (i.e., names) means that the applicable scale has more structure than that of the nominal one.

Ellerman (2014) glosses over this issue by focusing on how the T-accounts are combined. Since T-accounts are ordered pairs of numbers consisting of debit and credit entries, he simply shows by examples that to combine T-accounts one must add debits with debits and credits with credits. However, the first proposition addresses T-accounts, not the numbers within the T-accounts. The syntactic structure of the accounting language is such that words, the accounts' names, can be combined[14] and that the resulting words will bear an additional level of meaning that the users of accounting information can understand.

The second proposition states that the accounting equation is encoded by the group's neutral element, the 'zero T-account'. Several comments are required to explain what this means.

First, recall that zero is a special number for two reasons. When added to any number, that number does not change – this is why zero is the neutral element. Further, it helps define negative numbers: given a number, say 5, its inverse or negative is –5 because 5 + (–5) = 0.

Next, for the sake of measurement, consider being at some starting point which is associated with zero, and such that you may walk in either of two directions, say north or south along a road. If you walk 5 metres north and then 5 metres south, you end up at the origin as if you had not walked at all. This is a key feature of accounting measurement: accountants keep track of transactions in such a way that *by construction* the net effect is neutrality. The balance sheet always balances and the net flow of values is such that changes in assets are neutralised by changes in liabilities. We may now discuss the meaning of the second proposition.

---

[14] Not any two T-accounts may be *directly* combined, say inventory (an asset) with salaries (an expense). They may all be combined into the zero T-account, as explained next in this section. The point here is that there is more structure to the balance sheet and income statements than what is provided by the algebraic group alone.

One consequence of the above comments is that the group structure supports the T-accounts' expressing either stocks or flows. Ellerman (2014) summarises the point thus:

> *Each transaction is encoded as two or more T-accounts that add to the zero-account (double entry principle). Zero* [an earlier total stock] *added to zero* [a net flow] *equals zero* [the later total stock].[15]

Not all accountants may be aware that T-accounts also express flows. All of them understand that the journal is a list of double-entry transactions. However, this understanding means nothing other than zero T-accounts' expressing a flow of value. Given both propositions, one should accept that T-accounts can be combined at any point in time to express a stock of wealth *and* that they can also be combined over a time interval to express a flow of value.

As argued in Section 2, the users of accounting information want to address the two issues simultaneously: how entities combine economic resources and stakeholders' demands for the sake of measuring the entities' wealth, and also how entities keep track of economic processes for the sake of evaluating their performance. The two propositions herein inform that this is made possible by DEB further to DEB's satisfying the axioms of an algebraic group. Ellerman's (2014) comments and examples elaborate on this understanding by focusing on wealth. Willett's (1991) analysis of production structures elaborates on the same understanding by focusing on processes. We are not comparing both views; rather, we are arguing that both views stem from the same structure and are complementary in addressing the needs of users of accounting information. Thus, the focus on syntax is a relevant topic of investigation.

We still need to make sense of 'zero plus zero equals zero'. The second proposition refers to the set of *all* T-accounts. It informs that *total* assets equal *total* liabilities and that the *net* flow of values across assets and liabilities is neutral. This entails a taxonomic structure[16] for the set of accounts, as we explain next by means of an example.

---

[15] For a rigorous formalisation of this statement which sounds rather dull, the mathematically inclined reader is referred to Cruz Rambaud et al. (2010, ch.3).
[16] See Kay (1971). Our understanding is that the names of T-accounts qualify as a folk taxonomy.

*4.1 Illustration*

To account for an entity's stocks and flows under the double-entry procedure, one starts by representing the entity with the zero T-account **0** = (0, 0). However, there are many ways to convey this account, say (5, 5) or (123, 123); any T-account (x, x) expresses the zero element. Ellerman (2014) goes to great lengths to explain equivalence classes such that T-accounts may be seen as 'additive fractions' in the same way that 5/10 or 54/108 are fractions pertaining to the equivalence class defined by ½. Thus, if it is known that the entity has $1,234,567.89 in cash as its sole holdings, the two initial steps to encode information for accounting purposes are as follows. First, refer this amount as the basis for all subsequent information; it stands for 100% of assets and liabilities. Next, express the zero T-account in terms of the applicable accounting equation / balance sheet, thus: **0** = (100, 100) = (100, 0)$_{cash}$ + (0, 100)$_{liabilities}$.

Information at this level of aggregation is unlikely to be very useful. If on further inspection one finds out that the entity owes $493,827.16 to trade suppliers and the same amount to banks while its capital is $246,913.58, the next step is to break down the zero T-account further such that **0** = (100, 0)$_{cash}$ + (0, ⅖)$_{suppliers}$ + (0, ⅖)$_{banks}$ + (0, ⅕)$_{capital}$.

It should be clear that this procedure of refining information by means of partitions is limitless, at least in theory. In practice, a cost-benefit effect will trigger that determines where the process stops. In any case, and for the sake of our structural review herein, the result is a set of T-accounts endowed with a quotient structure that consists of *m* assets and *n* liabilities. There is no need for *m* to equal *n*, as we proceed to explain next.

So far the illustration is concerned with determining how the entity's wealth is organised. This is the stock-oriented view. To introduce the flow-oriented view, assume management intends to pay the suppliers using the available cash and to buy a machine using the long-term credit provided by the bank. This information is encoded as follows:

**0** = (⅕, 0)$_{cash1}$ + (⅖, 0)$_{cash2}$ + (⅖, 0)$_{cash3}$ + (0, ⅖)$_{suppliers}$ + (0, ⅖)$_{banks}$ + (0, ⅕)$_{capital}$

Accountants do not report in the financial statements how they intend to use the available cash. Management accountants, however, do work with budgets wherein the above allocations are encoded and acted upon. The point is that the structure of T-accounts is sufficiently rich to capture accountants' classifying resources according to how these resources are used. Indeed, a budget can be organised as a combined pro-forma balance sheet and income statement.

Further, Willett (1991) has formalised this view by suggesting that assets be conceived as unfinished activities. In terms of this example, his suggestion implies the zero T-account is separated as $\mathbf{0} = (\frac{1}{5}, 0)_{cash1} + (0, \frac{1}{5})_{capital} + (\frac{2}{5}, 0)_{cash2} + (0, \frac{2}{5})_{suppliers} + (\frac{2}{5}, 0)_{cash3} + (0, \frac{2}{5})_{banks}$. If so, when suppliers are paid, that portion of cash allocated to suppliers is a completed activity, and an asset, namely 'cash2', is taken out of the balance sheet. Furthermore, the machine being bought, 'cash3' reclassifies as 'machine': $\mathbf{0} = (\frac{1}{5}, 0)_{cash1} + (0, \frac{1}{5})_{capital} + (\frac{2}{5}, 0)_{machine} + (0, \frac{2}{5})_{banks}$. In turn, the machine is now partitioned in view of the productive activity it is expected to complete over time, say in five years it will contribute to costs in the exact amount of interest required by the bank. This is one instance of the matching principle: with the given data, this yields $\mathbf{0} = (2/25, 0)_{machine1} + (0, 2/25)_{interest1} + \ldots$ years 2, 3, 4 $+ (2/25, 0)_{machine5} + (0, 2/25)_{interest5}$. Thus, after the first year a portion of the machine is taken out of the balance sheet[17] in accordance to the activity being complete, that is, paying interest to the bank.

This example illustrates how the two propositions by Ellerman (2014) do indeed provide the structure used by accountants. This structure is not straightforwardly intuitive because, for instance, here the machine is likely to be considered from a physical perspective whereby it is an indivisible asset unit. To go beyond the intuition, one recognises that a machine worth $500,000 is, from an accounting perspective, a collection of 500,000 units of $1 such that each of those $1 units can be matched to $1 of future demands by stakeholders – for instance, here the requirement to pay interest to the bank over five years' time.

---

[17] In the usual accounting parlance, this is done by recording the proper amount as 'accumulated depreciation'.

We trust the example also illustrates the taxonomic structure of DEB. T-accounts' names obtain and convey wealth further to partitioning the zero T-account at some point in time. That partitioning results from assets being matched to the perceived liabilities in terms of activities.

## 5. *Discussion*

The main idea herein is the importance of time for measurement purposes. There is a need for a flow-oriented perspective in accounting measurement. We interpret the algebraic structure of DEB by Ellerman (2014) and argue that the T-account is the mathematical object to convey the accountants' view that the value of present resources, the assets, match the value of future demands by stakeholders, the liabilities – which include the shareholders' goal to make a profit. In short, the flow perspective is inherently embedded within the structure.

A conceptual consequence thereof is that assets cannot/should not be defined independently of liabilities. Already the current standard (i.e., by the IFRS as of March 2018) defines an asset as a present economic resource, whereas an economic resource is predicated on future benefits. A symmetrically similar definition applies to liabilities. Despite a liability being a present obligation, we understand that said obligation will be settled in the future. We believe the definitions should better emphasise the connection between present and future, and this can be accomplished by choosing the 'primacy of the T-account' over the 'asset/liability' view.

That choice would have implications for measurement. The standards allow for various measurement bases, each of which admits different levels of subjectivity when completing the measurement process. By building upon the primacy of the T-accounts, one would impose a consistency constraint on the results of any of the preferred bases, since a particular asset would not be measured independently from some other, properly matched liability. Further, this matching is at the basis of ratio analyses, as illustrated by the above example. Accounting numbers convey meaning in relative terms only: a million dollar profit is considered to be acceptable or not depending on the capital invested.

In terms of future developments, focusing on T-accounts and their built-in matching feature is likely to help frame managerial issues concerned with decision-making. We refer specifically to those issues arising from the analyses of costs and prices. Already costs (e.g., life-cycle costing) tend to be predicated on the future rather than on the past. The present framework wherein the zero T-account is partitioned in relevant activities addresses, at its most fundamental level, the concept of strategies and how to manage them. Indeed, the strategy concept is that which connects present resources to future goals while mapping how different but concomitant steps combine coherently towards achieving intermediate and final goals.

This paper is just a first step in that direction. Subsequent steps would require extending the algebraic structure to address vectors and how to deal with them in the classroom. Indeed, they are present in Ellerman's (2014) discussion. We do not engage with vectors here because the editors have instructed us otherwise. On the one hand, we have to limit ourselves to a partial response to that earlier paper. On the other hand, we have to remain within a limited discussion scope and thus we cannot address educational issues.

However, in response to an anonymous referee for whom this paper, along with Ellerman's, makes DEB look more complicated than it is, we wish to stress the following: Our paper is not intended to address the classroom teaching of accounting. It is intended to interest programme designers who must integrate different subjects within a single degree or qualification. Given the current nature of our modern and globalised economy, one would expect current accounting graduates to be able to go beyond bookkeeping and deal with complex issues of measurement, AI simulations, and strategic analyses.

## 6. Conclusion

When framing information, accountants consider both wealth (stocks of value) and income (flows of value). They report that information by means of two documents, the balance sheet and the income statement. Ellerman (2014) shows that both documents can be subsumed under a single mathematical object, the 'zero T-account', which encodes the accounting equation. As such, it aggregates both stock and flow information. The work of the accountant is to break down the stock of value into interesting and relevant categories (i.e., the several assets and liabilities) while simultaneously breaking down the flows of values into interesting and relevant categories (i.e., the several production processes and the several business strategies). Clearly, the accountant's dealing with both stocks and flows must be mutually consistent. The algebraic structure reviewed in this paper provides a first step towards framing this consistency issue.